\documentclass[preprint,preprintnumbers,amsmath,amssymb]{revtex4-1}
 \usepackage{graphicx}
\usepackage{latexsym}
\usepackage{natbib}
\usepackage{hyperref}
\usepackage{amsmath} 
\usepackage{appendix}

\usepackage{enumerate}
\usepackage{url}

\usepackage{color}
\usepackage{soul}

\begin{document}

\title{How does public opinion become extreme?}

\author{Marlon Ramos$^{1,2*}$, Jia Shao$^{3*}$, Saulo D. S. Reis$^{1,4*}$,
Celia Anteneodo$^{2}$, Jos\'e~S.~Andrade,~Jr $^{4}$, Shlomo Havlin$^5$,  Hern\'an A. Makse$^{1,4}$}

\affiliation{ $^1$ Levich Institute and Physics Department, City
  College of New York, New York, NY 10031, USA \\ $^2$ Departamento de
  F\'{\i}sica, PUC-Rio, 22451-900 Rio de Janeiro, Brazil\\ $^3$
  Bloomberg LP, New York, NY 10022, USA\\ $^4$ Departamento de
  F\'{i}sica, Universidade Federal do Cear\'{a}, 60451-970 Fortaleza,
  Cear\'{a}, Brazil \\ $^5$ Minerva Center and Physics Department,
  Bar-Ilan University, Ramat Gan 52900, Israel\\ 
	$^*$ These authors contributed equally to this work 
	}

\begin{abstract}

{\bf We investigate the emergence of extreme opinion trends in society
  by employing statistical physics modeling and analysis on polls that
  inquire about a wide range of issues such as religion, economics,
  politics, abortion, extramarital sex, books, movies, and electoral
  vote.  The surveys lay out a clear indicator of the rise of extreme
  views.  The precursor is a nonlinear relation between the fraction
  of individuals holding a certain extreme view and the fraction of
  individuals that includes also moderates, e.g., in politics, those
  who are ``very conservative'' versus ``moderate to very
  conservative'' ones.  We propose an activation model of opinion
  dynamics with interaction rules based on the existence of individual
  ``stubbornness'' that mimics empirical observations.  According to
  our modeling, the onset of nonlinearity can be associated to an
  abrupt bootstrap-percolation transition with cascades of extreme
  views through society.
Therefore, it represents an early-warning signal to forecast the
transition from moderate to extreme views.
Moreover, by means of a phase diagram we can classify societies
according to the percolative regime they belong to, in terms of
critical fractions of extremists and people's ties.}
 
\end{abstract}
\maketitle


The root causes of the rise of extreme opinions in society constitute
nowadays a matter of intense debate among leading scholars
\cite{hoffer,castellano:2009,klandermans,watts,review,Helbing}.  Over
the past few decades, there seems to be a worldwide trend towards the
division of public opinions about several issues, e.g., political
views, immigration, biotechnology applications, global warming, gun
control, abortion, LGBT rights amongst many others.  In many topics, a
marked dwindling of moderate voices is found with the concomitant
rising of extreme opinions \cite{Abramowitz:2008a,Layman:a,merkl}.
Not only in politics but also in simple topics such as books, movies,
fashion and other cultural topics, extreme positions sprout and the
opinion or attitude of an initially small group could become the rule.

How these tendencies settle in society is still a mystery. The degree
of social and economic development, the religious beliefs, the full
history, and many other factors, undoubtedly, all contribute to mold
the distribution of opinions of the members of a society. But,
besides those social features contributing to a collective mood,
interactions between individuals play also an important, often
underestimated, role. In the social network defined by the ties
between individuals, information, rumors, ideas, all travel. In this
process, new opinions can take form and existing ones can be either
strengthened or weakened. But to what extent does this interaction
help to shape the public opinion? Can extreme views arise just from
the interactions between individuals? The answers to these questions
can help us to understand the dynamics of polarization of public
opinions and make it possible to detect the trend to polarization.

Direct longitudinal statistics on the time evolution of individual
opinions at the large scale are hard to obtain. Fortunately, large
transverse data on the distribution of opinions of individuals about
different particular issues are publicly available from surveys.
These data have been obtained through polls, each one inquiring a
broad sample of people about their attitude towards a specific
subject, and offer a valuable evidence on the complex nature of public
opinions. The responses are usually categorized into attitudes, e.g.,
very favorable, somewhat favorable, somewhat unfavorable and very
unfavorable. For our purposes, people having very favorable or very
unfavorable opinions can be defined as those holding (either positive
or negative) extreme opinions.
 
The crux of the matter is to understand the dynamics of public opinion
from the available transverse data in order to forecast the trend to
polarization before it actually occurs \cite{earlywarning}.
From the analysis of these static data, we will extract clear evidence
of radicalization in groups in the form of nonlinear behaviour near
critical points and avalanche dynamics in belief spreading via
critical transitions in the bootstrap percolation universality
class. Such transitions shed light on the precise instance of the
transition when groups adopt more extreme views. Remarkably, these
transitions appear in a diversity of issues, indicating that the
results could be a generic feature of human opinion dynamics.

The central finding of our paper is the discovery of a sharp
statistical predictor of the rise of extreme opinion trends in society
in terms of a nonlinear behaviour of the number of individuals holding
a certain extreme view and the number of individuals with a moderate
opinion and extreme opinion. We analyze polls embracing a wide range
of issues such as religion, economics, politics, abortion,
extramarital sex, the electoral vote, and opinion on everyday consumer
products like books and movies. The surveys lay out a remarkable
nonlinear predictor of the rise of extreme opinion views.  This
predictor is ubiquitous across the diversity of polls and surveyed
countries, reflecting a remarkable generic feature of human opinion
dynamics.

The nonlinear methodology signals a tipping point at which a society
becomes extreme and has not been used before to predict opinion
trends, as far as we know.  The meaning of this nonlinearity is as
follows.  In general, for a statistical physics system of
non-interacting agents, isometry is expected. This means that the
system is extensive and the observables scale linearly with the system
size \cite{stat}. That is, if we double the number of particle, the
energy doubles as well, for instance. In term of our social system of
interest, a linear non-interactive extensive system implies that the
number of extreme people should scale linearly with the number of
people holding positive opinions. Thus, linearity is the byproduct of
non-interactions among the agent. On the other hand, it is well known
in statistical physics \cite{stat}, that correlations among the units,
that appear specially near a phase transition, lead to nonlinear
behaviour and non-extensivity. This effect is also called allometry in
the field of socio-physics and is currently being investigated, for
instance, in the scaling with the size of cities of different urban
indicators like technology activity \cite{bettencourt} and health
indicators \cite{hygor,erneson}. For instance, it is found that the
number of homicides scales superlinearly with the population of cities
while the number of suicides scales sublinearly; both cases being
examples of nonlinear allometric behaviour \cite{bettencourt,hygor}.

The onset of nonlinear behavior represents an early-warning signal
forecasting an abrupt critical transition from moderate to extreme
opinions, before it actually occurs.  The nonlinear behavior, which
anticipates an abrupt change, is easily detectable in society via
surveys and it measures the status of societies in the path towards
predominance of extreme attitudes.  By means of physical modeling, we
find that the nonlinearity forecasts the onset of cascades of extreme
view dissemination caused by the stubbornness of individuals.  We show
that the cascading is a consequence of an underlying
bootstrap-percolation transition occurring at the tipping point when
societies abruptly change from moderate to extreme.

\section*{Results}

{\bf Empirical findings.} To illustrate the polls, we consider a
typical survey from the Pew Research Center (see \hyperref[sec:reg]{Methods}).  
Participants from a given country are asked whether
they i) strongly believe, ii) believe, iii) disbelieve, or iv)
strongly disbelieve that religion is an important part of their
lives. 
Using these data, we first compute the fraction $f_e$ of people
holding an extreme view out of the total surveyed population in a
given survey and country. That is, we compute $f_{e}=N_e^+/N$ (or
$N_e^-/N$), where $N_e^+$ (or $N_e^-$) is the number of people
expressing an extreme positive view (or a negative one), and $N$ is
the total surveyed population. We then calculate the fraction of
people holding moderate to extreme views: $f=N^+/N$ (or $N^-/N$),
where $N^+$ (or $N^-$) is the number of individuals believing and
strongly believing in religion (or disbelieving and strongly
disbelieving).

Figure~\ref{fig:real_poll}a displays $f_e$ vs $f$, where each data
point represents the result of the survey carried out in a given
country and year.  The set of points, although spread, are neatly
correlated and follow a defined trend.  To extract the main
relationship between $f_e$ and $f$ without predetermined functional
form, we use nonlocal regression LOESS~\cite{rloess} with span $h=0.8$
as well as the Nadaraya-Watson estimator~\cite{Nadaraya:1964} (see
Sec. \hyperref[sec:reg]{Methods} for details).
The regression is represented by the solid line in
Fig.~\ref{fig:real_poll}a.  The result is paradigmatic of the
nontrivial dependency of $f_{e}$ on $f$ that defines the early-warning
signal at which a society starts to become extreme. For a relatively
small fraction of extremists, $f_e$ is approximately proportional to
$f$ (dotted straight line in Fig.~\ref{fig:real_poll}a).  This linear
behavior can be interpreted as arising from a system of
non-interacting individuals who form their opinions independently from
each other.  In the absence of interactions among people, the linear
regime would extend up to $f=1$.  However, at $f_e \approx 0.20$, a
noticeable departure from linearity is observed.  A nonlinear behavior
ensues, marking the onset of a surplus of extremists in comparison
with the expected number in the linear (non-interactive and extensive)
case.

A typical case study of transition towards extreme views is the
opinion about the economic situation 
after the European sovereign debt crisis of 2009.  The time evolution
of ($f,f_e$) for France, Italy, Greece, and Spain in
Fig.~\ref{fig:real_poll}b shows that
nonlinear behavior emerges after 2009, indicating that extreme
negativism has prevailed across societies. This result supports the
hypothesis that the departure from linearity marks the rise of extreme
views.

The observed nonlinearity is not a prerogative of religious or economic
issues where opinions frequently appear to be polarized, but extends
to many kinds of polls across the globe.  Polls ranging from
abortion to immigration (see details in \hyperref[sec:reg]{Methods})
are presented in Fig.~\ref{fig:real_poll}, all displaying similar
features.  It is a surprisingly ubiquitous behavior also found on much
simpler issues such as opinions on books and movies
(Fig.~\ref{fig:real_poll}~n-o).  Although the precise shape of
($f,f_e$)-curves changes from one poll to another, there seems to be a
universal trend very different to that found, for instance, in
shuffled data (Fig.~\ref{fig:real_poll}p). 

In Fig.~\ref{fig:real_poll}q, we show the results for 
state deputies, each point corresponds to one city for which we compute the fractions
of votes within each political orientation, as done for the other polls. 
This is a remarkable counter-example. We find a
dispersion pattern in $(f,f_e)$ similar to that which would appear if
people had chosen the political orientation of the candidates (from
extreme left to extreme right) in a random fashion. Indeed, the voting
data appear to be uncorrelated in similar way as that obtained in the
randomized data on books, Fig.~\ref{fig:real_poll}p.  Further research
is needed to reveal whether the absence of a trend in the Brazilian
electoral vote is a generic feature of elections at large.

In what follows we interpret the nonlinear behavior in terms of an
underlying critical transition from moderate to extreme views taking
place in society. Remarkably, the departure from linear behavior,
which appears for moderate $f_e$, forecasts a critical point marking
the precise transition from moderate behavior to extreme views.
Consequently, the ($f, f_e$)-curve, which can be easily obtained from
polls, readily predicts the onset of extreme opinion before the actual
transition has been materialized.

{\bf Modeling extreme opinion dynamics.} The features observed here cannot be 
explained by existing opinion models, 
as far as we know.
Most of them lead to consensus of a single opinion or to equal
fractions of opinions
\cite{Sznajd:2000a,Galam:2002a,Krapivsky:2003a}. Other ones allow
coexistence of minority and majority opinions
\cite{shao:2009a,biswas,CA:2012,singh2013}, but are not suitable to
describe the empirical data where we need to distinguish extreme from
moderate opinions.  All these models may constitute a sufficient
simplification to tackle certain problems, but they are not suitable
to study the emergence of extremisms where we need to distinguish
extreme from moderate opinions.  There are also the so-called {\em
  bounded confidence}
models~\cite{Deffuant:2000a,Hegselmann:2002a,castellano:2009} that
assume that only people with sufficiently close attitudes interact.
These models have been considered to study extreme opinion dynamics,
but lead to discontinuous distributions of opinions.

These observations call for a comprehensive simple
model to capture the underlying microscopic origin of extreme opinion
formation.  We propose a network model where the opinion of an
individual, $q$, takes real values between $-1$ and $+1$. Extreme
opinion is considered for $|q|>q_e$ and positive (negative) opinion
starts for $q>0$ ($q<0$). Without loosing generality, we consider
$q_e=0.50$ motivated by the four questions of most polls.
 
We introduce a parameter $a$ ($0\le a \le 1$) which gauges the
stubbornness of individuals, a realistic ingredient that we show to be
crucial to understand the nonlinear behavior in opinion spreading. 
The dynamics considers the previous opinion of the individual as well as
the average opinion $\bar{q}$ of the neighbors in the network
according to (see also Fig.~\ref{fig:rules}):

\begin{enumerate}[{\it (i)}]
\item $q\to\bar{q}$, if $|\bar{q}| > |q|$ and $q$ has the same sign as
  $\bar{q}$.
\item $q\to q$, if $(1-a)|q|\le|\bar{q}|\le|q|$ and $q$ has the same sign as
  $\bar{q}$.
\item $q \to \bar{q}+aq$, if [$\bar{q}<(1-a)q$ and $q > 0$] or
  [$\bar{q}>(1-a)q$ and $q<0$].
\end{enumerate}

Rule {\it (i)} determines that a node will adopt
the average opinion of its neighbors if this average is more extreme
that the node's opinion.  In fact, it is sound that people with a weak
opinion will be more likely influenced by people with a stronger one.
Notice that, even if the stubbornness parameter does not participate
explicitly in this rule, a subject who has a stronger opinion than its
contacts results to be more inflexible, since it is more difficult to
change its opinion. According to rule {\it (ii)}, no changes occur for
a range of intermediate opinions, this range being wider the larger
the stubbornness and the more stronger the node's opinion.
Finally, rule {\it (iii)} determines that, when the average opinion of
the neighbors is either opposite to or much less extreme than the
node's opinion, then the new opinion is $\bar{q}+aq$. That is, the new
opinion is determined not only by friends, but also partially by its
own opinion, weighted by the stubbornness $a$.  Thus, the role of
stubbornness $a$ is twofold: if $a$ is large, not only $\bar{q}$
should be farther enough from $q$ in order to change the node's
opinion, but $a$ also reduces the relative effect of its neighborhood.
In the limiting case $a=0$, the inflexibility range collapses
mimicking the most flexible individuals, easily influenced by the
close environment and assuming the average value of the neighbors,
similarly to majority rule models~\cite{Galam:2002a}.  

Stubbornness is a crucial ingredient to have an heterogeneous
population with different opinions.  In the absence of stubbornness
($a=0$), all the three rules reduce to the single prescription of
adopting the average value of the neighbors, yielding consensus of a
single opinion as in the majority rule model of \cite{Galam:2002a}.
Differently, when setting $a>0$, people with initially different
opinions will not be easily convinced and heterogeneity of opinions
will persist in the final state, yielding a continuous probability density 
function of opinions. 

We simulate the model on an Erd\"os-R\'enyi (ER) network, a general
class of random networks with a Poisson degree distribution and with
the small-world property \cite{BookCaldarelli}, with average degree
$\langle k \rangle$, starting with $f_0$ fraction of nodes with
positive opinion (we set $a=1$ in all simulations).  
To define the initial state of the opinion
dynamics on top of a chosen network of size $N$, we select $f_0$ that
gives the initial fraction of nodes with positive opinion.  After
that, we  select $f_0 N$ nodes and assign to each one of them
a random opinion value $q$ uniformly distributed between $0$ and $+1$.
To the remaining $(1-f_0)N$ nodes, we assign a random value of $q$
uniformly distributed between $-1$ and 0.  Then, at each time step $t$
the opinions $q$ of all nodes in the network are synchronously updated
according to the rules defined above.  Positive extremists are a
minority for any initial condition.  We then compute the fractions $f$
and $f_e$ in the final state controlled by $f_0$.  As shown in
Fig.~\ref{fig:comp}a, the model reproduces very well $(f,f_e)$ of
religion data.

 {\bf Phases of extreme opinion.} 
Next, we discuss how the phenomenology of the
model allows us to interpret the nonlinearity in terms of changes in
the microscopic dynamics of beliefs spreading. These changes are
expressed in well-defined transitions between the different phases of
the final state depicted in Fig.~\ref{fig:rules}c. The transitions
from one phase to another are characterized by the percolative
behavior of extremists and their networks of contacts.  The behavior
of the connected components of extremists (named {\em e-clusters},
Fig.~\ref{fig:rules}c) reveals the origin of the nonlinearity.
Changing $f_0$, the system passes through three distinct phases
separated by two critical transition points as exemplified in
Fig.~\ref{fig:rules}c.  The phenomenology of the transitions 
is closely related to activation models like bootstrap
percolation~\cite{chalupa:1979a,goltsev,baxter:2010a,baxter:2011}, the
opinion model of Watts~\cite{watts,granovetter} and the
multi-percolation model of competition of innovations of Helbing {\it
  et al.}  \cite{Helbing:2012}. Indeed, there is a correspondence
between the dynamics of vertex activation in bootstrap
percolation~\cite{chalupa:1979a,baxter:2010a} and the change from
moderate to extreme opinions ({\em e-activation}) in our model both
starting from an initial configuration where nodes are active with
probability $f_0^e=(1-q_e)f_0$ (see below).

The purpose of the model is then to interpret the nonlinear behavior
in terms of critical phase transitions which cannot be directly
measured from real data since the contact network of ties is usually
unknown at the large scale. The model identifies the following phases:

{\it  Moderate Phase I:} For low $f_0$, we observe small isolated
e-clusters. The size of the largest e-cluster, $s_1^e$, as a function
of $f_0$ vanishes (Fig. \ref{fig:measures}) and the behavior of $(f,
f_e)$ remains approximately linear.

{\it Incipient Phase II.} Above a critical value, $f_{0c_1}$,
a giant e-component of size $s_1^e$ emerges which occupies a
non-vanishing fraction of the network (Figs.~\ref{fig:measures}a and
\ref{fig:measures}e).  The critical point $f_{0c_1}$ is also signaled
by the peak in the size of the second largest e-cluster, $s_2^e$. The
order of this transition is determined by $\langle k \rangle$ in
comparison with a critical value $k_c=4.5\pm0.1$.  For $\langle k
\rangle > k_c$, $s_1^e$ (Fig.~\ref{fig:measures}a), as well as $f_e$
and $f$ (Fig.~\ref{fig:measures}c), present a discontinuity at
$f_{0c_1}$; a fingerprint of an abrupt first-order transition. For
$\langle k \rangle < k_c$, the transition is second order like in
ordinary percolation. The size $s_1^e$ increases continuously at
$f_{0c_1}$, $s_2^e$ presents a peak (Fig.~\ref{fig:measures}e), and
$f_e$ and $f$ also increase smoothly (Fig.~\ref{fig:measures}g).

After a giant e-cluster appears, a collective phenomenon in avalanches
of extreme opinion spreading takes place.  We quantify the avalanche
dynamics inspired by similar dynamics appearing in bootstrap
percolation \cite{baxter:2010a,baxter:2011}.

In bootstrap percolation 
~\cite{baxter:2010a,baxter:2011} nodes in a given network can take
two values, active or inactive.  At the beginning of the dynamics, a
fraction $f_a$ of nodes chosen at random are set into the active
state, the rest are inactive. An inactive node becomes active only if
it has at least $k$ active neighbors, where $k$ is a fixed parameter
of the model, while active nodes remain in this state forever.  The
activation rule is iteratively applied until the system reaches a
final state with no further changes.
A variant has been introduced by Watts~\cite{watts} in which the
activation condition is given by a minimal fraction of active
neighbors, instead of a minimal fixed number of neighbors $k$.

In bootstrap percolation, when a giant cluster of active site exists,
an infinitesimal change of the fraction of active nodes can trigger an
avalanche of activations.  This cascade process is related to the
existence of sub-critical clusters of activatable nodes.
A node belongs to a subcritical cluster if its number of active
neighbors external to the cluster is one less that the threshold
degree necessary for activation~\cite{baxter:2010a}. 
When a sub-critical node gains an active neighbor, it becomes active
and, as a consequence, its connected neighbors in the cluster in turn
gain a new active neighbor, and a cascade occurs. In contrast, in our
case, the activation rules are far more complex to allow a clear
definition of sub-critical nodes.
 In fact, the e-activation itself of a vulnerable node does not
guarantee the activation of its activatable nearest neighbors.
Furthermore, indirect activation is also possible: a node $i$ might
be transitively activated through some already activated
intermediary, as soon as the node $i$ receives an extra contribution
to its $\bar{q}$ due to the modification of one of its nearest
neighbors.

 In order to detect and characterize the possible avalanches, we
circumvented that difficulty by perturbing the system. 
We choose a node with opinion $0<q<q_e=0.5$ and substitute it by
$q=1$, measuring the number of vulnerable nodes $S$ that become
extremist in the new stable state.  
Figure ~\ref{fig:rules}c  shows the result of the process described above. 
 
We accumulate data for all nodes with opinions below $q_e$ (triggered
one at a time) that succeeded in triggering an avalanche and repeat
for several realizations.
The average size of the avalanches $\langle S \rangle$ and the largest
avalanche size $S^{^\ast}$ were computed as a function of $f_0$. 
We find that $S$ is small around
$f_{0c_1}$ but increases rapidly with $f_0$. The largest
avalanche size $S^\ast$ as a function of $f$ is plotted in
Fig.~\ref{fig:comp}a. It indicates that the nonlinear trend in ($f,
f_e$) in model and empirical data is accompanied by the increase of
avalanche sizes.
Thus, we associate the onset of the nonlinear regime in the incipient
extreme phase where the system starts to be susceptible to changes,
and small perturbations can generate a cascade of extreme opinion
spreading.

{\it Extreme Phase III.} $S^\ast$ peaks at a second transition point
$f_{0 c_2}$ (Figs. \ref{fig:measures}b and \ref{fig:measures}f)
signaling the transition to a phase where the whole society has become
extreme.  This transition can be smooth or abrupt according to
$\langle k \rangle$. If $\langle k \rangle > k_c$, the transition is
sharp and first-order. The distribution of avalanche size develops a
power-law tail with scaling exponent $3/2$ (inset
Fig.~\ref{fig:measures}b).  The value of this critical exponent
suggests that the model is in the universality class of bootstrap
percolation ~\cite{baxter:2010a,baxter:2011,Helbing:2012}.  
Furthermore, the activation dynamics in bootstrap
percolation~\cite{baxter:2010a} and the opinion model of
Watts~\cite{watts} exhibit hybrid transitions as in our model: a
combination of a jump (as in first order transitions) and a power law
(as in second order transitions) near the critical point.
Close to the critical point, the size of the largest e-cluster behaves like
\begin{equation}
|s^e_1-s^e_{1c}|\sim|f_0-f_{0c}|^{\zeta},
\end{equation}
where $f_{0c}$ refers to either $f_{0c_1}$ and $f_{0c_2}$, and with
the exponent $\zeta \approx 1/2$ (see Fig.~\ref{fig:htr}), 
like in bootstrap percolation~\cite{baxter:2010a,baxter:2011}. 
We notice that these are
hybrid transitions, and the approach to the critical point in terms of
power laws is given from above and below for $f_{0c_1}$ and
$f_{0c_2}$, respectively.  This result further suggests that our
model, although not the same as bootstrap percolation, could be in the
same universality class.

The sharp peak of $S^\ast$ (Fig. \ref{fig:measures}b) reflects the
discontinuity in $s_1^e$ at $f_{0c_2}$ (Fig.~\ref{fig:measures}a),
which is also seen in $f$ and $f_e$ (Fig.~\ref{fig:measures}c).  After
this abrupt jump, almost all nodes belong to the giant e-cluster. When
$\langle k \rangle < k_c$, $S^\ast$ presents a smeared peak at $f_{0
  c_2}$ (Fig. \ref{fig:measures}f). The $3/2$ power-law decay found
for $\langle k \rangle> k_c$ applies approximately to the envelope of
the distributions of avalanche sizes (inset
Fig.~\ref{fig:measures}f). In this case, the approach to the extreme
phase is progressive in terms of $f$ and $f_e$
(Figs. ~\ref{fig:measures}g and \ref{fig:measures}h).
 
The impact of this critical scenario on $(f, f_e)$ is illustrated in
Figs.~\ref{fig:comp}a, ~\ref{fig:measures}d and
~\ref{fig:measures}h. They show that the onset of nonlinearity in the
Incipient Phase II is associated to the increase of cascade
sizes.  The origin of nonlinearity is the presence of cascades of
extremists in phase II and the onset of nonlinearity is a predictor of
more drastic changes that occur when the size of the avalanches
becomes maximal.

The different phases predicted by the model are
represented in Fig.~\ref{fig:comp}b into a phase diagram defined in
terms of precise critical values of $f_e$ and $\langle k\rangle$. 
It displays the line of percolation
transition separating moderate and incipient extreme phases predicted
by the model, whose main trait is the absence and presence of a giant
e-cluster, respectively, and the transition to the extreme phase.  In
the case of religion polls, we find $\langle k \rangle = 4.2$ in
Fig. \ref{fig:comp}b which is obtained by fitting the data $(f_e,f)$
in Fig. \ref{fig:comp}a using all the data points from all the
countries. Once the value of $\langle k \rangle$ is obtained, then we
can plot the particular countries in the phase diagram since we also
know exactly the value of $f_e$.

By means of the interpretation provided by the model, 
we classify societies according to their extreme level; the phase
diagram measures the status of societies in the path towards
predominance of extreme attitudes. Selected data from 
religion polls from Fig.~\ref{fig:comp}a are projected onto the phase
diagram, Fig.~\ref{fig:comp}b. Most of the countries are located in
Phase II and a few are in Phase III, where the majority of the
population has become extreme.
For instance, we find that, in terms of positive opinion about how
religion is important in peoples life, Pakistan and Tunisia are in
Phase III, while Brazil is at the transition point between Phase II
and III. USA is also very close to the transition point closely
approaching Phase III. We notice that the position of a country in the
phase diagram can be changed by an increase or decrease of either
$f_e$ or, more importantly, $\langle k \rangle$. The effective degree
can be easily increased by the use of social media, for example. Thus,
for instance, USA might enter Phase III in religion attitude by just
increasing its effective degree from its current $\langle k
\rangle=4.2 $ to $\langle k \rangle = 5$. This would produce a first
order abrupt transition to Phase III.  Other countries like Mexico,
Italy and Japan are in the incipient Phase II.  Finally, China belongs
to the moderate Phase I in terms of positive attitudes towards religion.

This classification may have important implications, since we could
detect whether a country is at the edge of an abrupt change to extreme
phase produced either by an increase of $f_{e}$ or $\langle k\rangle$
(for instance, by increasing connectivity by the use of social media).

\section*{Final remarks}
 
A natural situation for extreme behavior is human opinion as studied
here. The consistency between real data and model predictions is
suggestive of a possible broader scope of the present statistical
analysis. This good agreement makes it a candidate for predictor of
other aspects of human collective behavior involving beliefs and
decision-making where opinion cascades prevail~\cite{watts}, such as
competition of market innovations~\cite{Helbing:2012,shiller}.  For
instance, the nonlinear early-signature might be able to anticipate
wide adoption of consumer products, as soon as the nonlinearity
appears in consumer ratings of items such as books and movies. Further
research is planned to investigate the applicability of nonlinear
analysis to human collective behavior at large.

\section*{Methods}
\label{sec:reg}

{\bf Nonparametric regression.} We consider nonparametric regression procedures  to obtain
a smooth set of points from each
set of scattered data  $(x_i, y_i)$, $i = 1,..., n$, as those in Fig.
\ref{fig:real_poll}:
the locally weighted regression (LOESS) and the Nadaraya-Watson (NW)
regression.

We used LOESS, with span $h = 0.8 $ to extract the main trend of
$(f,f_e)$ as well as the NW estimator.
%
\\[2mm]  \noindent
{\it LOESS: }
the estimated values $\hat{y_i}$  for each point $x_i$ are obtained
through a weighted least-squares fitting procedure~\cite{rloess}.
A weight function $W$ that depends on the distance $h_i$ to the $r$th
nearest neighbor of point $i$ is used.
The $k = 1,..., n$, (with $k\neq i$) weights for each point $x_i$ are
given by
\begin{equation}
w_k(x_i)=W\left( \frac{x_k-x_i}{h_i}\right),
\label{eq:weight}
\end{equation}
where $W$ is the tricubic weight function
$$ W=\left\{\begin{array}{cc}
(1-|x|^3)^3&\mbox{, if}\quad |x|<1\\
0 &\mbox{, if}\quad |x| \geq 1 \,.
\end{array}\right.
$$
Equation (\ref{eq:weight}) determines the estimated $\hat{y_i}$
in reference \cite{rloess}.
%
\\[2mm] \noindent
{\it Nadaraya-Watson:} we  construct the kernel smoother
function~\cite{Nadaraya:1964}
\begin{equation}
\hat{m}_h(x)=\frac{\sum_i^n K_h(x-x_i)Y_i}{\sum_i^n K_h(x-y_i)}\,,
\label{eq:estimator}
\end{equation}
where $K_h(x - x_i)$ is a Gaussian kernel of the form,
\begin{equation}
\label{eq:kernel}
K_h(x-x_i)=\exp\left[ \frac{\left(x-x_i\right)^2}{2h^2} \right]\,,
\end{equation}
with bandwidth $h$ estimated by least squares cross-validation method. \\

{\bf Description of polls used in Fig.~\ref{fig:real_poll} }
We provide information about the data used in each
panel depicted in Fig.~\ref{fig:real_poll}.
For the survey data, obtained for example from the Pew Research
Center, we present explicitly, when available, {\it (i)} the question
used in each survey, {\it (ii)} the original URL where the data can be
found, {\it (iii)} the number of countries where the surveys were
performed, {\it (iv)} the number of surveys performed which is larger
than the number of countries in {\it (iii)} since the surveys are
performed over many years for a given country, and {\it (v)} the dates
when the surveys were performed.\\

\noindent
{\bf a.} Religion: \\
Question: How important is religion in your life -- very important,
somewhat important, not too important, or not at all important?   \\
Source: Pew Research Center  \\
URL: \url{http://www.pewglobal.org/question-search/?qid=408&cntIDs=&stdIDs=}\\
Total number of countries: 59\\
Total number of surveys: 231\\ 
Surveys date: Summer 2002, Spring 2005, Spring 2006, Spring 2007,
Spring 2008, Spring 2009, Fall 2009, Spring 2010, Spring 2011,
Late Spring 2011, and Spring 2012.\\

\noindent
{\bf b.} Economic situation:  \\
Question: Now thinking about our economic situation, how would you describe
the current economic situation in (survey country) - is it very good,
somewhat good, somewhat bad or very bad?  \\
Source: Pew Research Center   \\
URL: \url{http://www.pewglobal.org/question-search/?qid=753&cntIDs=&stdIDs=}\\
Total number of countries: 59\\
Total number of surveys: 260\\ 
Surveys date: Summer 2002, Spring 2007, Spring 2008, Spring 2009,
Fall 2009, Spring 2010, Spring 2011, Late Spring 2011, Spring 2012,
and Spring 2013.\\
The time evolution is shown for the following cases: \\
Countries: France, Italy, Greece, and Spain.\\
Total number of surveys: 24\\ 
Surveys date: Summer 2002 (France and Italy), Spring 2007 (France, Italy, and Spain),
Spring 2008 (France and Spain), Fall 2009 (France, Italy, and Spain),
Spring 2009 (France and Spain), Spring 2010 (France and Spain),
Spring 2011 (France and Spain), Spring 2012 (France, Italy, Greece, and Spain),
and Spring 2013 (France, Italy, Greece, and Spain).\\

\noindent
{\bf c.} Jews:\\
Question: Please tell me if you have a very favorable, somewhat favorable,
somewhat unfavorable, or very unfavorable opinion of Jews  \\
Source: Pew Research Center  \\
URL: \url{http://www.pewglobal.org/question-search/?qid=834&cntIDs=&stdIDs=}\\
Total number of countries: 32\\
Total number of surveys: 131\\ 
Surveys date: Spring 2004, Spring 2005, Spring 2006, Spring 2008,
Spring 2009, Spring 2010, Spring 2011, and Late Spring 2011.\\

\noindent
{\bf d.} Muslims:\\
Question: Please tell me if you have a very favorable, somewhat favorable,
somewhat unfavorable, or very unfavorable opinion of Muslims   \\
Source: Pew Research Center  \\
URL: \url{http://www.pewglobal.org/question-search/?qid=836&cntIDs=&stdIDs=}\\
Total number of countries: 32\\
Total number of surveys: 135\\ 
Surveys date: Spring 2004, Spring 2005, Spring 2006, Spring 2008, Spring 2009,
Spring 2010, Spring 2011, and Late Spring 2011.\\

\noindent
{\bf e.} Christians:\\
Question: Please tell me if you have a very favorable, somewhat favorable,
somewhat unfavorable, or very unfavorable opinion of Christians  \\
Source: Pew Research Center \\
URL: \url{http://www.pewglobal.org/question-search/?qid=828&cntIDs=&stdIDs=}\\
Total number of countries: 32\\
Total number of surveys: 133\\ 
Surveys date: Spring 2004, Spring 2005, Spring 2006, Spring 2008,
Spring 2009, Spring 2010, Spring 2011, and Late Spring 2011.\\

\noindent
{\bf f.} Business ties: \\
Question: What do you think about the growing trade and business ties between
(survey country) and other countries - do you think it is a very good thing,
somewhat good, somewhat bad or a very bad thing for our country?  \\
Source: Pew Research Center \\
URL: \url{http://www.pewglobal.org/question-search/?qid=1011&cntIDs=&stdIDs=}\\
Total number of countries: 55\\
Total number of surveys: 184\\
Surveys date: Summer 2002, Spring 2007, Spring 2008, Spring 2009,
Spring 2010, Spring 2011, and Late Spring 2011.\\

\noindent
{\bf g.}  Immigration:  \\
Question: As I read another list of statements, for each one, please tell me
whether you completely agree, mostly agree, mostly disagree or completely
disagree with it... We should restrict and control entry of people into our
country more than we do now.  \\
Source: Pew Research Center  \\
URL: \url{http://www.pewglobal.org/question-search/?qid=54&cntIDs=&stdIDs=}\\
Total number of countries: 54\\
Total number of surveys: 128\\ 
Surveys date: Summer 2002, Spring 2007, Spring 2009, and Fall 2009.\\

\noindent
{\bf h.} United States: \\
Please tell me if you have a very favorable, somewhat favorable, somewhat
unfavorable, or very unfavorable opinion of the United States.  \\
Source: Pew Research Center   \\
URL: \url{http://www.pewglobal.org/question-search/?qid=844&cntIDs=&stdIDs=}\\
Total number of countries: 59\\
Total number of surveys: 351\\ 
Surveys date: Summer 2002, March 2003, May 2003, Spring 2004,
Spring 2005, Spring 2006, Spring 2007, Spring 2008, Spring 2009,
Spring 2010, Spring 2011, Late Spring 2011, Spring 2012, and Spring 2013.\\

\noindent
{\bf i.} Foreign influence (protection against):  \\
Question: As I read another list of statements, for each one, please tell me
whether you completely agree, mostly agree, mostly disagree or completely
disagree with it... Our way of life needs to be protected against foreign influence. \\
Source:  Pew Research Center  \\
URL: \url{http://www.pewglobal.org/question-search/?qid=51&cntIDs=&stdIDs=}\\
Total number of countries: 52\\
Total number of surveys: 119\\ 
Surveys date: Summer 2002, Spring 2006, Spring 2007, Spring 2009,
and Spring 2012.\\

\noindent
{\bf j.} Success (determined by external forces): \\
Question: Please tell me whether you completely agree, mostly agree, mostly
disagree or completely disagree with the following statement... Success in
life is pretty much determined by forces outside our control  \\
Source:  Pew Research Center   \\
URL: \url{http://www.pewglobal.org/question-search/?qid=908&cntIDs=&stdIDs=}\\
Total number  of countries: 55\\
Total number of surveys: 155\\
Surveys date: Summer 2002, Spring 2007, Spring 2008,
Spring 2009, Fall 2009, Spring 2011, and Late Spring 2011.\\

\noindent
{\bf k.} Abortion:  \\ 
Question: Do you agree  very much, a little, not really, 
not at all with the statement... If a woman doesn't want children, 
she should be able to have an abortion.  \\
Source: Euro RSCG/TNS Sofres\\
URL: \url{http://en.wikipedia.org/wiki/Societal_attitudes_towards_abortion}\\
Total number of countries: 10 (European only)\\
Total number of surveys:  10\\
Surveys date: May 2005\\

\noindent
{\bf l.} Same-sex marriage: \\ Question: Please tell me whether you
strongly favor, favor, oppose, or strongly oppose it... Allowing gay
and lesbian couples to marry legally?\\ Source: Pew Research Center
\\ URL:
\\ \url{http://pt.scribd.com/doc/131666438/Polls-on-Attitudes-on-Homosexuality-Gay-Marriage}\\ Total
number of countries: 1\\ Total number of surveys: 28\\ Surveys date:
May 1996-October 2012\\
 
\noindent
{\bf m.} Extramarital sex: \\ Question: What about a married person
having sexual relations with someone other than the marriage partner,
it is always wrong, almost always wrong, wrong only sometimes, or not
wrong at all? \\ Source: NORC/GSS
\\ URL:\\ \url{http://pt.scribd.com/doc/131666438/Polls-on-Attitudes-on-Homosexuality-Gay-Marriage}\\ Total
number of countries: 1 (USA)\\ Total number of surveys: 23\\ Surveys
date: 1973-2010\\

\noindent
{\bf n.} IMDB Movies: \\ We collect ratings (from 1 to 10 stars) of
imdb.com movies with number of opinionators greater than 1,000. We
crawled all the votes until March 28, 2013. We exclude TV episodes.
Each datapoint in Fig. \ref{fig:real_poll}n is a movie out of the
19,405 total.  We convert the star ratings into opinion as
follows:\\ extreme positive opinion ($N_e^+$): 9 and 10
stars,\\ positive opinion ($N^+$): 7, 8, 9 and 10 stars,\\ negative
opinion ($N^-$): 1, 2, 3 and 4 stars,\\ extreme negative opinion
($N_e^-$): 1 and 2 stars.\\ URL:
\url{http://www.imdb.com/search/title?at=0&sort=release_date_us&title_type=feature,tv_movie,tv_series,tv_special,mini_series,documentary,game,short,video,unknown&user_rating=1.0,10}\\ Total
number of movies: 301,743; with more than 1,000 ratings: 19,405 \\

\noindent
{\bf o.} Amazon Books: \\ We collected ratings, from 1 to 5 stars, of
books at sale on amazon.com with a minimum of 50 opinionators.  Each
datapoint in Fig. \ref{fig:real_poll}o is a book out of the total of
16,390.  We convert the star ratings into opinion as
follows:\\ extreme positive opinion ($N_e^+$): 5 stars,\\ positive
opinion ($N^+$): 4 and 5 stars,\\ negative opinion ($N^-$): 1 and 2
stars,\\ extreme negative opinion ($N_e^-$): 1 star.  \\ URL:
\url{http://www.amazon.com/}\\ Total number of books: 291,428; with
more than 50 ratings: 16,390 \\

\noindent
{\bf p.} Amazon Books (shuffled): \\ For each book on Amazon presented
in Fig~\ref{fig:real_poll}o we randomly redistributed the positive
votes (4 and 5 stars) and the negative ones (1 and 2 stars),
separately.\\

\noindent
{\bf q. Brazilian elections (state deputies in 2010):} \\ At the 2010
electoral dispute, there were 27 eligible parties in Brazil: PMDB, PT,
PP, PSDB, PDT, PTB, PTdoB, DEM, PR, PSB, PPS, PSC, PCdoB, PV, PRB,
PRP, PMN, PSL, PTC, PSDC, PHS, PTN, PRTB, PSOL, PSTU, PCB, and PCO. We
obtain the political orientation for each one of these parties from \\
\url{http://en.wikipedia.org/wiki/List_of_political_parties_in_Brazil}\\
(accessed on 11/22/2013):
\begin{itemize}
\item Extreme-left: PSTU, PCB, PCO (Total: 3)
\item Left: PT, PSB, PCdoB, PSOL (Total: 4)
\item Center-left: PSDB, PDT, PTB, PPS, PV, PMN (Total: 6)
\item Center: PMDB, PTdoB, PRB, PRP, PSL, PHS, PTN, PRTB (Total: 8)
\item Center-right: PTC, PSDC (Total: 2)
\item Right: PP, DEM, PR, PSC (Total: 4)
\end{itemize}

We analyze the 2010 Brazilian election for state deputies, which
correspond to state legislative assemblies representatives
(Fig.~\ref{fig:real_poll}q). 
These data are available at
\url{http://agencia.tse.jus.br/estatistica/sead/odsele/votacao_partido_munzona/votacao_partido_munzona_2010.zip}.

For each city in Brazil, we compute the number of votes received by
the parties associated to each one of the six political
orientations. Note that there is no extreme-right party in
Brazil. Arbitrarily, we take votes on extreme-left, left, and
center-left parties as negative opinion, $N^-$. The votes on the
center, center-right, and right parties are considered as positive
opinion, $N^+$. We consider as extreme opinions the votes on
extreme-left and left parties, $N_e^-$, and the votes on center-right
and right parties, $N_e^+$, respectively. This choice is motivated by
the fact that very small fractions of the electorate correspond to
orientations of extreme-left and center-right.
Currently, there are 32 parties in
Brazil, where 5 new parties were created in the country since the
2010's election. None of the parties in 2010 considered in the present
work was dissolved.


\section*{Acknowledgements} 

This work was supported by NSF and ARL under Cooperative Agreement
Number W911NF-09-2-0053. Additional financial support was provided by
LINC and Multiplex EU projects, and the Brazilian agencies CNPq,
CAPES, Faperj and FUNCAP. We thank S. Alarc\'on for useful
discussions.

\section*{Author Contributions}
H.A.M designed research, M.R, J.S., S.D.S.R, C.A., J.S.A. Jr., S.H. and
H.A.M. performed research, and wrote the manuscript.

\section*{Additional information}
Competing financial interests: The authors declare no competing financial interests.


\clearpage

\begin{figure*}[h!]  
\includegraphics[width=.75\textwidth]{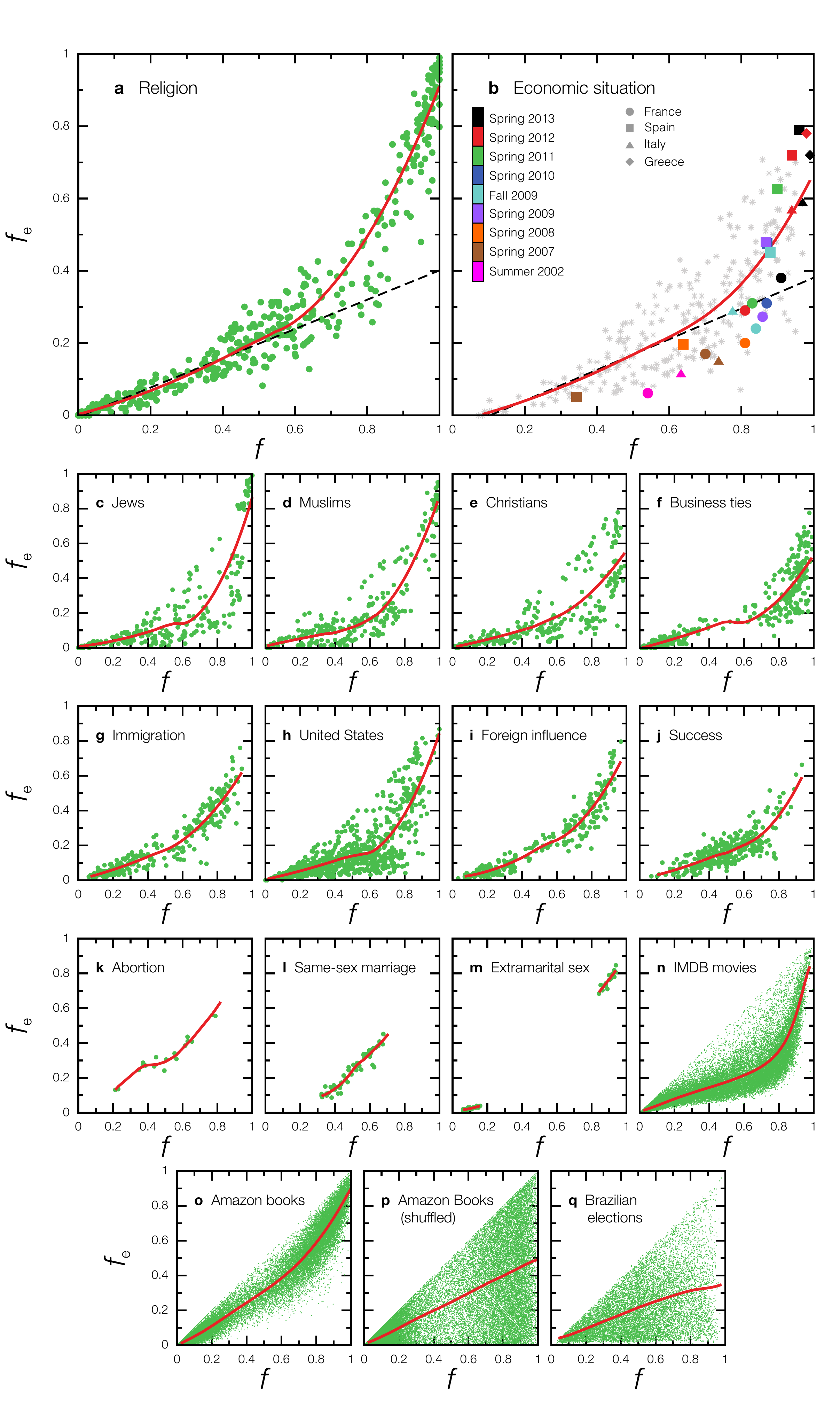}	
\caption{}
\label{fig:real_poll}
\end{figure*}

FIG. \ref{fig:real_poll}. {\bf Empirical observations}. Dependence of
the fraction of extremists $f_e$ on the fraction $f$ of all people
sharing an opinion, obtained from the outcomes of polls inquiring
about a wide spectrum of issues as explained in Sec. \hyperref[sec:reg]{Methods}.  
For instance, in ({\bf a}) participants
from a given country are asked whether they {\it (i)} strongly
believe, {\it (ii)} believe, {\it (iii)} disbelieve, or {\it (iv)}
strongly disbelieve that religion is an important part of their lives.
For each country and/or year, $f$ and $f_e$ were computed as explained
in the text, for both favorable and unfavorable responses.  The solid
line is a nonparametric regression (see Methods).  In ({\bf a}) and
({\bf b}), the dotted line depicts the linear behavior expected for a
non-interactive group. In ({\bf b}), the fractions for unfavorable
responses in surveys inquiring about the feeling on the economic
situation are plotted. The time evolution (in color scale) of
($f,f_e$) is depicted for France, Italy, Greece, and Spain. A
nonlinear behavior emerged in these countries after the European
sovereign debt crisis of 2009.

\begin{figure*}[h!]
\includegraphics[width=0.9\textwidth]{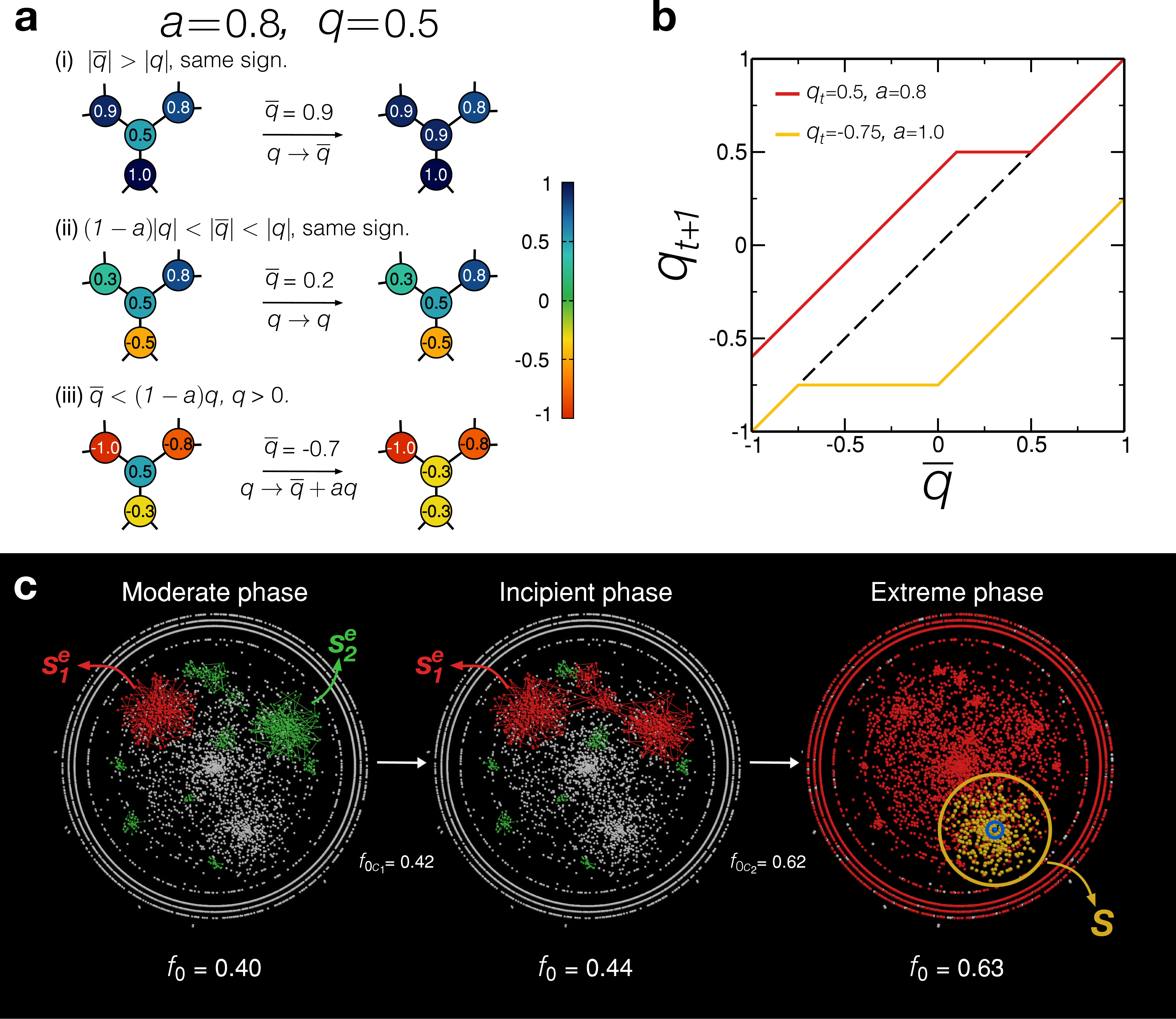} 
\caption{}
\label{fig:rules}
\end{figure*}

FIG. \ref{fig:rules}. {\bf Opinion model.}  ({\bf a}) Consider a node
with degree 3 which holds a moderate opinion $q=0.5$, and stubbornness
$a=0.8$.  There are three possible situations in the model rules
according to $\bar{q}$: {\it (i)} If $\bar{q}(=0.9) > q$ and of the
same sign as $q$, the node's opinion becomes more extreme, $q \to
\bar{q}$.  {\it (ii)} If $\bar{q} < q$ but larger than a fraction of
$q$ given by $1-a$, then the nearest neighbors cannot change the
opinion of the node due to its stubbornness. Consequently, the opinion
remains the same.  {\it (iii)} When the average opinion of the nearest
neighbors is more moderate or opposite in sign (as in the panel), it
can influence the node's opinion.  Since in this case $\bar{q}= -0.7$,
the positive opinion of the node changes, becoming $q= -0.3$.  ({\bf
  b}) Diagram showing the new opinion of a node at step $t+1$,
$q_{t+1}$, as a function of the average opinion of the node's nearest
neighbors at step $t$, $\bar{q}$.  Two typical cases are depicted.
Red curve: with moderate positive opinion $q_{t}=0.5$ (blue) and
moderate stubbornness $a=0.8$ (this case corresponds to ({\bf a})).
Orange curve: with extreme negative opinion $q_{t}=-0.75$ and
stubbornness $a=1$.  The larger the value of $a$ the wider the
inflexibility range of rule {\it (ii)} where opinion does not change.
({\bf c}) Illustration of the different phases for different values of
$f_0$ in a typical ER network of size $N=10,000$ and $\langle k
\rangle=4$.  In Moderate Phase I, extremist clusters are mostly
isolated. The largest e-cluster is in red and top ten in green, white
nodes are moderate, most of the activity is concentrated in the
3-core, and the concentric circles are the 1 and 2-shells
\cite{kitsak,sen1,sen2}.  In Incipient Phase II an incipient giant e-cluster
first appears (red). The system is increasingly more susceptible to
perturbations; the yellow cluster in Extreme Phase III depicts a
cascade resulting by the change to extreme of a single node in blue.
Deep in the Extreme Phase III most nodes (in red) have become
extremists.

\begin{figure*}[h!]
\includegraphics[width=0.9\textwidth]{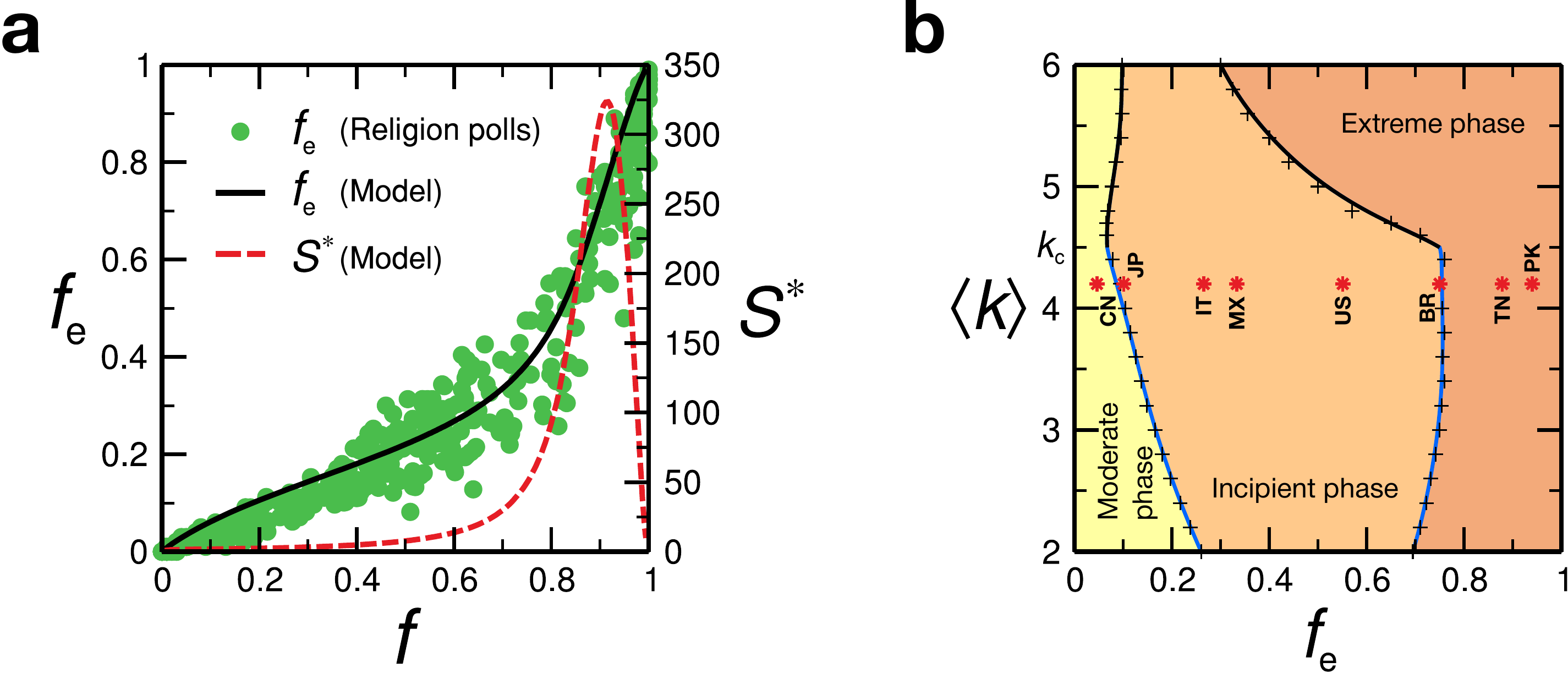}
\caption{}
\label{fig:comp}
\end{figure*}

FIG. \ref{fig:comp}. {\bf Model and poll data.} ({\bf a}) $f_e$ vs $f$
for the religion polls and fitting using the model with $N=10^4$,
$a=1$ and $\langle k \rangle = 4.2$.  The model closely matches the
empirical result.  We plot the largest avalanche size $S^\ast$
obtained by damaging the network as explained in the text. The onset
of nonlinear behavior and cascading avalanches coincide.  ({\bf b})
Extreme phase diagram from modeling in terms of $f_e$ and $\langle k
\rangle$. The transition lines separating the three phases at
$f_{ec_1}$ and $f_{ec_2}$ are analogous to $f_{0c_1}$ and $f_{0c_2}$,
respectively. Black lines correspond to first-order transitions for
$\langle k \rangle > k_c$, and blue lines correspond to continuous
transitions for $\langle k \rangle < k_c$.  Moderate Phase I: there is
no giant e-cluster.  Incipient Phase II: a giant e-cluster
appears, with increasing cascading effects. Extreme Phase III:
characterized by the consensus of extremists for sufficiently high
mean degree. The symbols represent selected countries from religion
polls in ({\bf a}) encompassing the whole spectrum of phases (names in
Internet two-letter code).  $\langle k \rangle$ should be interpreted
as the effective average degree through which opinion spreads rather
than the actual number of ties of the individuals which could be much
larger. The effective average degree is obtained from the fitting in
({\bf a}).

\begin{figure*}[h!]
\includegraphics[width=0.9\textwidth]{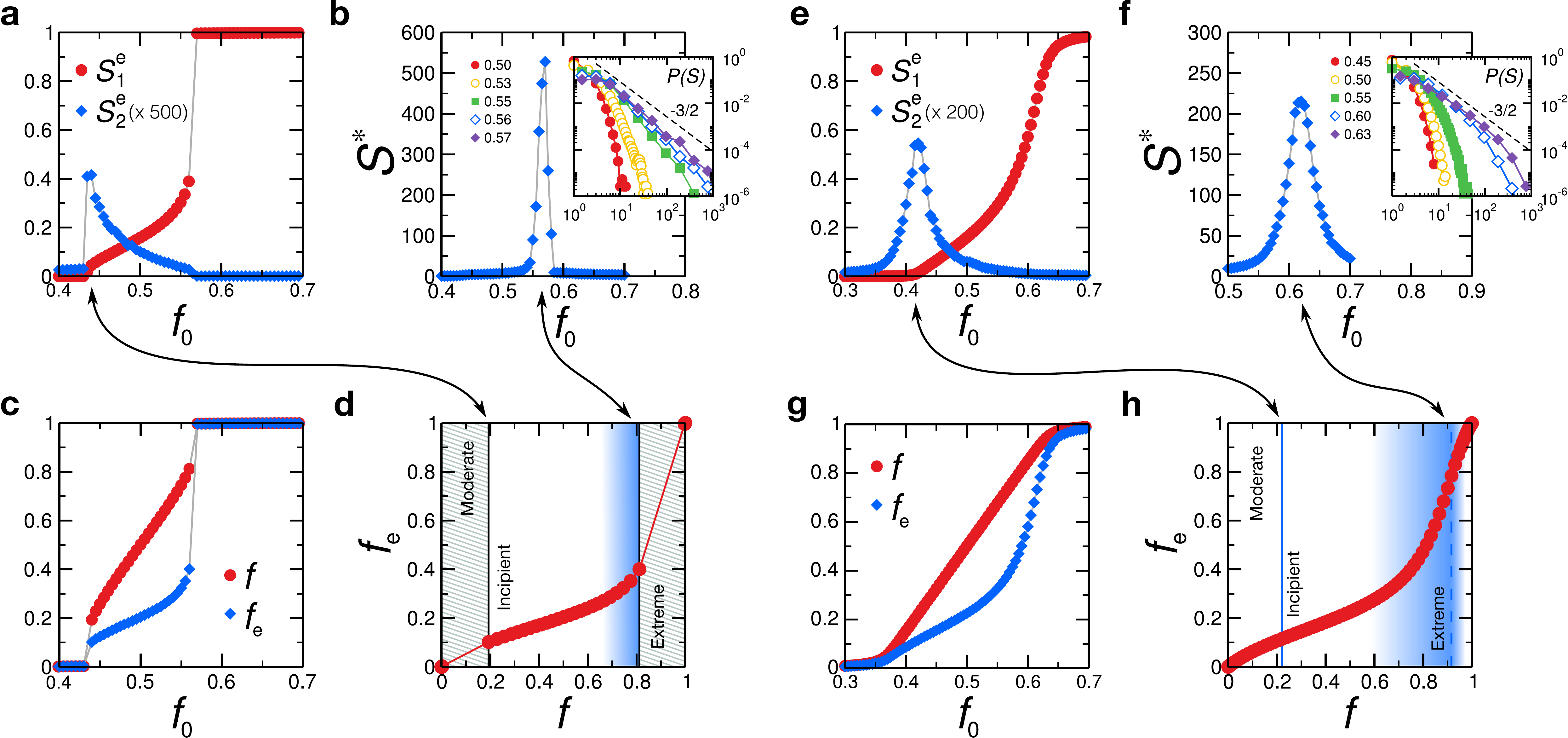}
\caption{}
\label{fig:measures}
\end{figure*}

FIG. \ref{fig:measures}. {\bf Critical transitions.} ({\bf a})-({\bf d})
For $\langle k \rangle = 5 > k_c$.  ({\bf e})-({\bf h}) For $\langle k
\rangle = 4 < k_c$. Displayed results are an average over 50 ER networks (except
for ({\bf b}) and ({\bf f}) where we use 300 networks) and we set
$a=1$. ({\bf a}) and ({\bf e}) $s_1^e$ and $s_2^e$ vs $f_0$.  Cluster
sizes are normalized by the size of the network ($N=10^5$).  ({\bf b})
and ({\bf f}) Largest cascade size $S^\ast$ vs $f_0$.  The inset shows
the distribution of cascade sizes for different values of $f_0$,
exhibiting power-law scaling ($N=10^4$).  ({\bf c}) and ({\bf g})
$f_e$ and $f$ vs $f_0$ ($N=10^5$).  ({\bf d}) and ({\bf h}) Curves
$(f, f_e)$ to highlight the nonlinear behavior.
The hatched regions in ({\bf d}) correspond to the jumps in the
first-order transitions, hence inaccessible in the infinite size
limit. The bluish colored areas in ({\bf d}) and ({\bf h}) represent
the region of large cascading $S^\ast$ regime from ({\bf b}) and ({\bf
  f}), respectively. They show that the nonlinearity is associated
with the occurrence of progressively larger cascades as $f$ increases.

\begin{figure*}[h!]
\includegraphics[width=0.65\textwidth]{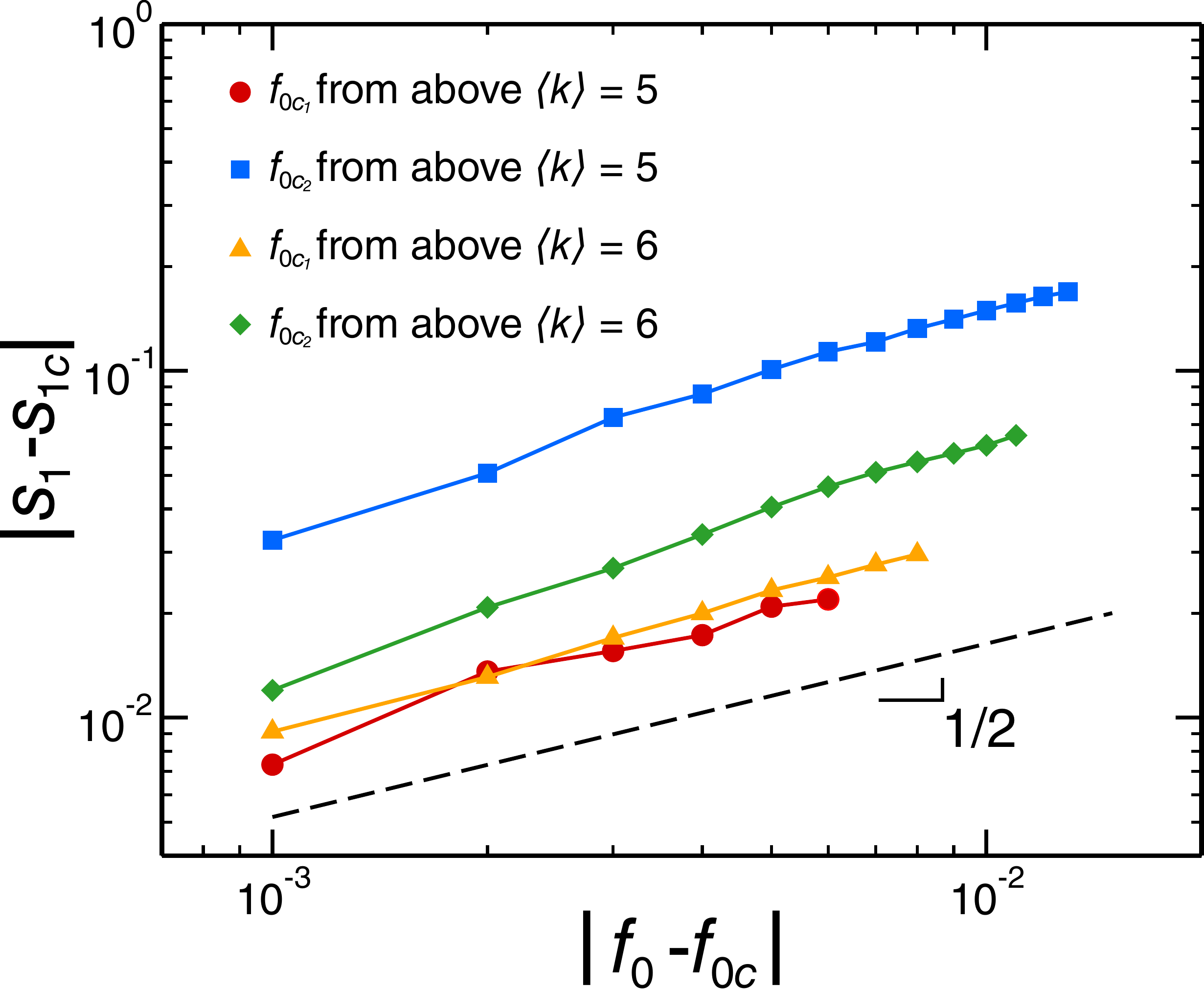}
\caption{}
\label{fig:htr}
\end{figure*}

FIG. \ref{fig:htr}. {\bf Scaling at the hybrid transition}, for two
cases of high connectivity: $\langle k\rangle=5$ and $\langle
k\rangle=6$, at the two critical points.  The exponent is close to
1/2, that of bootstrap percolation.  We used ER networks of size
$10^5$.

\clearpage

\clearpage

\clearpage

\clearpage

\clearpage

\end{document}